\documentclass[
 reprint,
amsmath,amssymb,
aps,
prl,
longbibliography,
citeautoscript
]{revtex4-1}

\usepackage[Symbol]{upgreek}
\usepackage{amssymb}
\usepackage{mathrsfs}
\usepackage{graphicx}
\usepackage{dcolumn}
\usepackage{bm}
\usepackage{graphicx}
\usepackage{dcolumn}
\usepackage{bm}

\usepackage[english]{babel}
\usepackage{hyperref}
\usepackage{times}
\usepackage{amsfonts,amssymb,bm}
\usepackage{amsmath}
\usepackage{epsfig}
\usepackage{mathrsfs}
\usepackage{color,soul}
\usepackage{bbold}
\usepackage{subfig}
\usepackage{pstricks}
\usepackage{mathtools}
\usepackage{textcomp}
\usepackage{braket}
\usepackage{empheq}
\usepackage{amsmath}
\usepackage{amsthm}
\usepackage[babel]{csquotes}
\usepackage[font=small]{caption}
\usepackage[labelfont=bf,labelsep=quad,justification=raggedright]{caption}
\usepackage{epstopdf}

\begin{document}

\preprint{APS/123-QED}

\title{Supertoroidal light pulses: Propagating electromagnetic skyrmions in free space}

\author{Yijie Shen$^{1,*}$, Yaonan Hou$^{1}$, Nikitas Papasimakis$^{1}$, and Nikolay I. Zheludev$^{1,2}$}
\affiliation{
	{$^{1}$Optoelectronics Research Centre \& Centre for Photonic Metamaterials, University of Southampton, Southampton SO17 1BJ, United Kingdom}\\
	{$^{2}$Centre for Disruptive Photonic Technologies, School of Physical and Mathematical Sciences and The Photonics Institute, Nanyang Technological University, Singapore 637378, Singapore}
}

\date{\today}
\begin{abstract}
\noindent Topologically complex transient electromagnetic fields give access to nontrivial light-matter interactions and provide additional degrees of freedom for information transfer. An important example of such electromagnetic excitations are space-time non-separable single-cycle pulses of toroidal topology, the exact solutions of Maxwell described by Hellwarth and Nouchi in 1996 and recently observed experimentally. Here we introduce a new family of electromagnetic excitation, the supertoroidal electromagnetic pulses, in which the Hellwarth-Nouchi pulse is just the simplest member. The supertoroidal pulses exhibit skyrmionic structure of the electromagnetic fields, multiple singularities in the Poynting vector maps and fractal-like distributions of energy backflow. They are of interest for transient light-matter interactions, ultrafast optics, spectroscopy, and toroidal electrodynamics.
\end{abstract}

\maketitle

\noindent\textbf{Introduction} -- 
Topology of complex electromagnetic fields is attracting growing interest of the photonics and electromagnetics communities~\cite{berry2000making,soskin2016singular,lu2014topological,khanikaev2017two,ozawa2019topological,mortensen2019topological}, while topologically structured light fields find applications in super-resolution microscopy~\cite{pu2021unlabeled,du2019deep}, metrology~\cite{yuan2019detecting,sederberg2020vectorized}, and beyond~\cite{shen2019optical,yao2011orbital,chen2019orbital}. For example, the vortex beam with twisted phase, akin to a Mobius strips in phase domain, can carry orbital angular momentum with tuneable topological charges enabling advanced optical tweezers, machining, and communication applications~\cite{shen2019optical,yao2011orbital,chen2019orbital}. The complex electromagnetic strip and knots structures were also proposed as information carriers~\cite{bauer2015observation,bauer2016optical,larocque2020optical}. Recently observed electromagnetic skyrmions are relevant to topological skyrmions quasiparticles in high-energy physics and condensed matter~\cite{tsesses2018optical,du2019deep,dai2020plasmonic,gao2020paraxial,cuevas2021optical}. They have sophisticated vector topology~\cite{fert2017magnetic,yu2010real,nagaosa2013topological}, and enable applications in super-resolution microscopy~\cite{du2019deep}, ultrafast imaging~\cite{davis2020ultrafast}, and give rise to new types of spin-orbit optical forces~\cite{halcrow2020attractive}. 

While a large body of work on topological properties of structured continuous light beams may be found in literatures, works on the topology of the time-dependent electromagnetic excitations and pulses only start to appear. For instance, the ``Flying Doughnut'' pulses, or toroidal light pulses (TLPs) first described in 1996 by Hellwarth and Nouchi~\cite{hellwarth1996focused}, with unique spatiotemporal topology predicted recently~\cite{zdagkas2019singularities}, have only very recently observed experimentally~\cite{zdagkas2021observation}. 

Fuelled by a combination of advances in ultrafast lasers or THz emitters and in our ability to control the spatiotemporal structure of light~\cite{papasimakis2018pulse,mcdonnell2021functional} together with the introduction of new experimental and theoretical pulse characterization methods~\cite{zdagkas2021spatio,shen2020measures}, TLPs are attracting growing attention. Indeed, TLPs exhibit their complex topological structure with vector singularities and interact with matter through coupling to toroidal and anapole localized
modes~\cite{papasimakis2016electromagnetic,raybould2017exciting,savinov2019optical}. Generation of TLPs in the optical and THz parts of the spectrum now paves the way towards new forms of spectroscopy, sensitive to toroidal and anapole excitations, and new information transfer schemes.

In this paper, we report that the Hellwarth and Nouchi pulses, are, in fact, the simplest example of an extended family of pulses that we will call supertoroidal light pulses (STLPs).  We will show that supertoroidal light pulses introduced here exhibit complex topological structures that can be controlled by a single numerical parameter. The STLP display skyrmion-like arrangements of the transient electromagnetic fields organized in a fractal-like, self-affine manner, while the Poynting vector of the pulses feature singularities linked to the multiple energy backflow zones.

\begin{figure*}[t!]
	\centering
	\includegraphics[width=\linewidth]{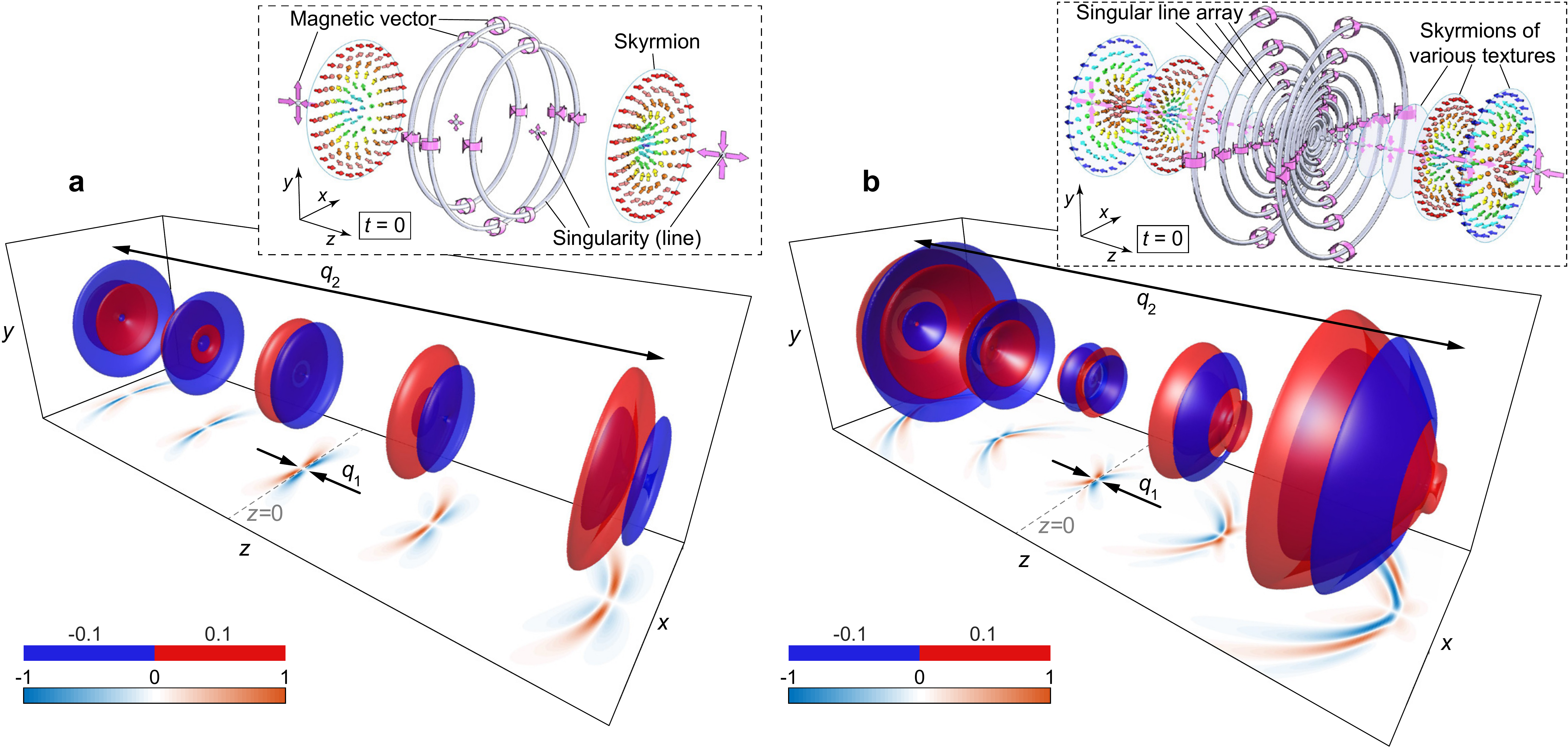}
	\caption{\textbf{From toroidal to supertoroidal light pulses:} \textbf{a,b}, 
	Isosurfaces for the electric fields of (\textbf{a}) the fundamental TLP $\text{Re}[E_\theta(\mathbf{r},t)]$, and (\textbf{b}) a STLP $\text{Re}[E_\theta^{(\alpha)}(\mathbf{r},t)]$ of $\alpha=5$, at amplitude levels of $E=\pm0.1$ and the Rayleigh range of $q_2=100q_1$, at different times of $t=0$, $\pm q_2/(4c)$, and $\pm q_2/(2c)$. $x$-$z$ cross-sections of the instantaneous electric field at $y=0$. The insets in (a) and (b) are schematics of spatial topological structures of magnetic vector fields at focus ($t=0$) for the fundamental TLP and STLP, respectively. The gray dots and rings mark the distribution of singularities in magnetic field, large purple arrows mark the direction of magnetic field vector, and the smaller coloured arrows show the skyrmionic structures in magnetic field.} 
	\label{f1}
\end{figure*}

\begin{figure*}[t!]
	\centering
	\includegraphics[width=0.96\linewidth]{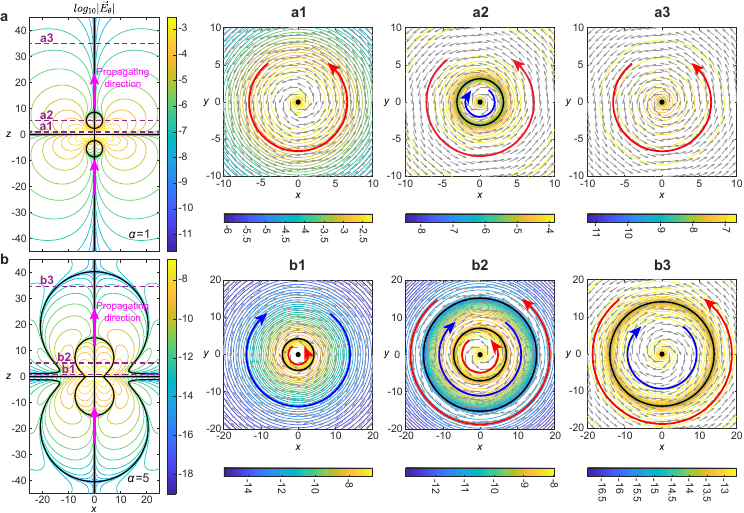}
	\caption{\textbf{Electric field topology of toroidal and supertoroidal light pulses:} \textbf{a,b}, The isoline plots of the electric field in the $x$-$z$ plane for ({a}) the fundamental TLP, $\text{Re}[{E}_{\theta}(\mathbf{r},t=0)]$, and ({b}) the STLP of $\alpha=5$, $\text{Re}[{E}^{(\alpha=5)}_{\theta}(\mathbf{r},t=0)]$, in logarithmic scale. Solid black lines indicate the zeros of the electric field. Dashed lines marked as {a1}-{a3} and {b1}-{b3} indicate the $z$-levels of the cross-sections on the right panels correspondingly. Panels \textbf{a1}-\textbf{a3} and \textbf{b1}-\textbf{b3} present isoline plots of the electric field and arrow plots of the electric field direction in the $x$-$y$ plane, in logarithmic scale. Solid black lines and black dots mark the points and areas, where the electric field vanishes. Blue and red arrows indicate the two opposite azimuthal directions of the electric fields. Unit for coordinates: $q_1$.} 
	\label{f2}
\end{figure*}

\vspace{0.2cm}
\noindent\textit{\textbf{Results}}

\noindent\textbf{Supertoroidal electromagnetic pulses} -- Following Ziolkowski, localized finite-energy pulses  can be obtained as superpositions of ``electromagnetic directed-energy pulse trains''~\cite{ziolkowski1989localized}. A special case of the localized finite-energy pulses was investigated by Hellwarth and Nouchi~\cite{hellwarth1996focused}, who found the closed-form expression describing a single-cycle finite energy electromagnetic excitation with toroidal topology obtained from a scalar generating function $f(\mathbf{r},t)$ that satisfies the wave equation $( \nabla^2-\frac{1}{{{c}^{2}}}\frac{{{\partial }^{2}}}{\partial {{t}^{2}}})f\left( \mathbf{r},t \right)=0$,
where $\mathbf{r}=(r,\theta,z)$ are cylindrical coordinates, $t$ is time, $c=1/\sqrt{\varepsilon_0\mu_0}$ is the speed of light, and the $\varepsilon_0$ and $\mu_0$ are the permittivity and permeability of medium. Then, the exact solution of $f(\mathbf{r},t)$ can be given by the modified power spectrum method~\cite{ziolkowski1989localized,hellwarth1996focused}, as $f(\mathbf{r},t)={f_0}/\left[{( {{q}_{1}}+i\tau){{( s+{{q}_{2}})}^{\alpha }}}\right]$, where $f_0$ is a normalizing constant, $s=r^2/(q_1+i\tau)-i\sigma$, $\tau=z-ct$, $\sigma=z+ct$, $q_1$ and $q_2$ are parameters with dimensions of length and act as effective wavelength and Rayleigh range under the paraxial limit, while $\alpha$ is a real dimensionless parameter that must satisfy $\alpha\ge1$ to ensure finite energy solutions. Next, transverse electric (TE) and transverse magnetic (TM) solutions are readily obtained by using Hertz potentials. The electromagnetic fields for the TE solution can be derived by the potential $\mathbf{A}(\mathbf{r},t)=\mu_0\bm{\nabla} \times\mathbf{\hat{z}}f(\mathbf{r},t)$ as $\mathbf{E}(\mathbf{r},t)=-{{\mu }_{0}}\frac{\partial }{\partial t}\bm{\nabla} \times \mathbf{A}$ and $\mathbf{H}(\mathbf{r},t)=\bm{\nabla} \times(\bm{\nabla}\times \mathbf{A})$ \cite{ziolkowski1989localized,hellwarth1996focused}. Finally assuming $\alpha=1$, the electromagnetic fields of the TLP are described by~\cite{hellwarth1996focused}:
\begin{align}
& {E_\theta}=-4i{{f}_{0}}\sqrt{\frac{{{\mu }_{0}}}{{{\varepsilon }_{0}}}}\frac{r(q_1+q_2-2ict)}{{{\left[ {{r}^{2}}+\left( {{q}_{1}}+i\tau \right)\left( {{q}_{2}}-i\sigma  \right) \right]}^3}}\label{Ef}\\
& {{H}_{r}}=4i{{f}_{0}}\frac{r({{q}_{2}}-{{q}_{1}}-2iz)}{{{\left[ {{r}^{2}}+\left( {{q}_{1}}+i\tau  \right)\left( {{q}_{2}}-i\sigma  \right) \right]}^3}}\\ 
& {{H}_{z}}=-4{{f}_{0}}\frac{{{r}^{2}}-\left( {{q}_{1}}+i\tau  \right)\left( {{q}_{2}}-i\sigma  \right)}{{{\left[ {{r}^{2}}+\left( {{q}_{1}}+i\tau  \right)\left( {{q}_{2}}-i\sigma  \right) \right]}^3}}\label{Hf}
\end{align}
where the electric field $E_\theta$ is azimuthally polarized with no longitudinal or radial components, whereas the magnetic field is oriented along the radial and longitudinal directions, $H_r$ and $H_z$, with no azimuthal component. Equations (\ref{Ef}-\ref{Hf}) derived by Hellwarth and Nouchi for $\alpha=1$ show the simplest example of TLPs.
Here we explore the general solution for values of $\alpha\ge1$. In the TE case, electric and magnetic fields are given by (see detailed derivation in Supplementary Information):
\begin{widetext}
\begin{align}
& E_{\theta }^{(\alpha )}=-2\alpha i{{f}_{0}}\sqrt{\frac{{{\mu }_{0}}}{{{\varepsilon }_{0}}}}\left\{ \frac{(\alpha +1)r{{({{q}_{1}}+i\tau )}^{\alpha -1}}({{q}_{1}}+{{q}_{2}}-2ict)}{{{\left[ {{r}^{2}}+\left( {{q}_{1}}+i\tau  \right)\left( {{q}_{2}}-i\sigma  \right) \right]}^{\alpha +2}}}-\frac{(\alpha -1)r{{({{q}_{1}}+i\tau )}^{\alpha -2}}}{{{\left[ {{r}^{2}}+\left( {{q}_{1}}+i\tau  \right)\left( {{q}_{2}}-i\sigma  \right) \right]}^{\alpha +1}}} \right\}\label{Egf}\\
& H_{r}^{(\alpha )}=2\alpha i{{f}_{0}}\left\{ \frac{(\alpha +1)r{{({{q}_{1}}+i\tau )}^{\alpha -1}}({{q}_{2}}-{{q}_{1}}-2iz)}{{{\left[ {{r}^{2}}+\left( {{q}_{1}}+i\tau  \right)\left( {{q}_{2}}-i\sigma  \right) \right]}^{\alpha +2}}}-\frac{(\alpha -1)r{{({{q}_{1}}+i\tau )}^{\alpha -2}}}{{{\left[ {{r}^{2}}+\left( {{q}_{1}}+i\tau  \right)\left( {{q}_{2}}-i\sigma  \right) \right]}^{\alpha +1}}} \right\}\label{hr} \\ 
& H_{z}^{(\alpha )}=-4\alpha {{f}_{0}}\left\{ \frac{{{({{q}_{1}}+i\tau )}^{\alpha -1}}\left[ {{r}^{2}}-\alpha \left( {{q}_{1}}+i\tau  \right)\left( {{q}_{2}}-i\sigma  \right) \right]}{{{\left[ {{r}^{2}}+\left( {{q}_{1}}+i\tau  \right)\left( {{q}_{2}}-i\sigma  \right) \right]}^{\alpha +2}}}+\frac{(\alpha -1){{({{q}_{1}}+i\tau )}^{\alpha -2}}\left( {{q}_{2}}-i\sigma  \right)}{{{\left[ {{r}^{2}}+\left( {{q}_{1}}+i\tau  \right)\left( {{q}_{2}}-i\sigma  \right) \right]}^{\alpha +1}}} \right\}  \label{Hgf}
\end{align}
\end{widetext}
For $\alpha=1$, the electromagnetic fields in Eqs.~(\ref{Egf}-\ref{Hgf}) are reduced to that of the fundamental TLP Eqs.~(\ref{Ef}-\ref{Hf}). Moreover, the real and imaginary parts of equations (\ref{Egf}-\ref{Hgf}), simultaneity fulfill Maxwell equations and therefore represent real electromagnetic pulses (see Supplementary Materials).

While propagating in free space toroidal and supertoroidal pulses exhibit self-focusing. Figure~\ref{f1} shows the evolution of the fundamental ($\alpha=1$) TLP and STLP ($\alpha=5$) upon propagation through the focal point. In the former case, the pulse is single-cycle at focus ($z=0$) becoming $1\frac{1}{2}$-cycle at the boundaries of Rayleigh range $z=\pm q_2/2$  (Fig.~\ref{f1}\textbf{a}).
On the other hand, STLPs with $\alpha>1$ (Fig.~\ref{f1}\textbf{b}, $\alpha=5$) exhibit a substantially more complex spatiotemporal evolution where the pulse is being reshaped multiple times upon propagation (See Video~1 and Video~2 in Supplementary Materials for the dynamic evolution of the fundamental TLP and STLPs). 

\begin{figure*}[t!]
	\centering
	\includegraphics[width=\linewidth]{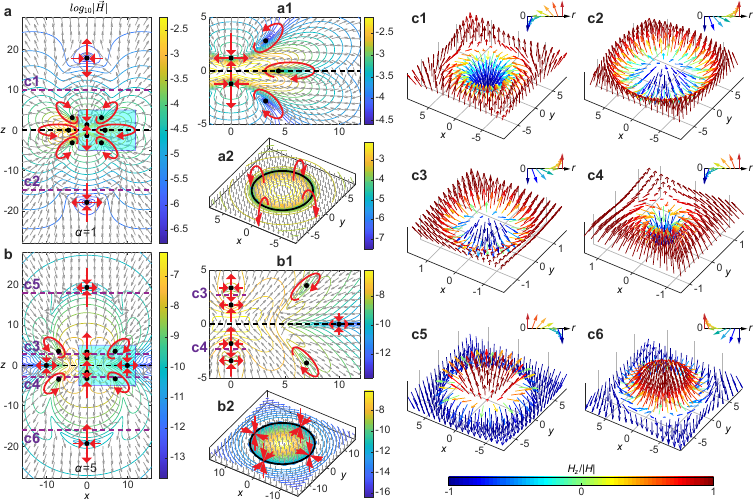}
	\caption{\textbf{Magnetic field topology of toroidal and supertoroidal light pulses:} \textbf{a,b}, Isoline and arrow plots of the magnetic fields in the $x$-$z$ plane for ({a}) the fundamental TLP and ({b}) the STLP of $\alpha=5$, in logarithmic scale. Black dots indicate the zeros of the magnetic field with red arrows correspondingly marking the saddle or vortex style of the vector singularities. Panels \textbf{a1} and \textbf{b1} present the zoom-ins of the regions highlighted by blue in (a) and (b) respectively. Panels \textbf{a2} and \textbf{b2} present isoline and arrow plots of the magnetic fields at $z=0$ planes, in logarithmic scale, marked by the black dashed lines in (a) and (b) respectively, where the magnetic fields vanish along the circular solid black lines with red arrows marking the styles of singularities (vortex for (a2) and saddle for (b2)). \textbf{Skyrmionic structures in magnetic fields of toroidal and supertoroidal light pulses:} \textbf{c}, Various textures of N\'eel-type skyrmionic structure observed at various transverse planes (see dashed purple lines in (a-b)) for the fundamental TLP (c1-c2) and the STLP of $\alpha=5$ (c3-c6), which are demonstrated by the arrows with color-labeled longitudinal component value of magnetic field. The up-right insert of each panel shows the basic texture of the skyrmionic structure. Unit for coordinates: $q_1$.} 
	\label{f3}
\end{figure*}

\begin{figure*}[t!]
	\centering
	\includegraphics[width=\linewidth]{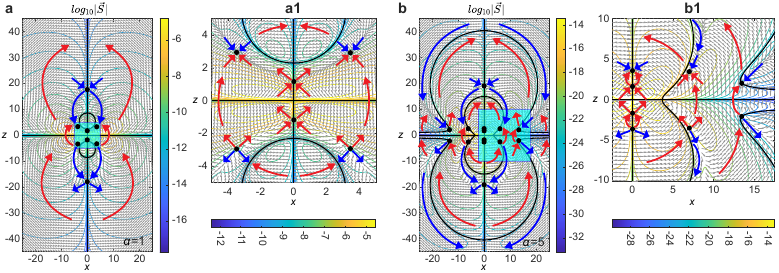}
	\caption{\textbf{Poynting vector topology of toroidal and supertoroidal light pulses:} \textbf{a,b}, Contour and arrow plots of the Poynting vector fields in the $x$-$z$ plane, in logarithmic scale, for (a) the fundamental TLP and (b) the STLP of $\alpha=5$. Panels (\textbf{a1}) and (\textbf{b1}) present zoom-ins of the areas highlighted by green in (a-b), respectively. Solid black lines and dots mark the zeros of the Poynting vector. Red and blue arrows indicate areas with forward and backward energy flow, respectively. Unit for coordinates: $q_1$.} 
	\label{f4}
\end{figure*}
\begin{figure*}[t!]
	\centering
	\includegraphics[width=1\linewidth]{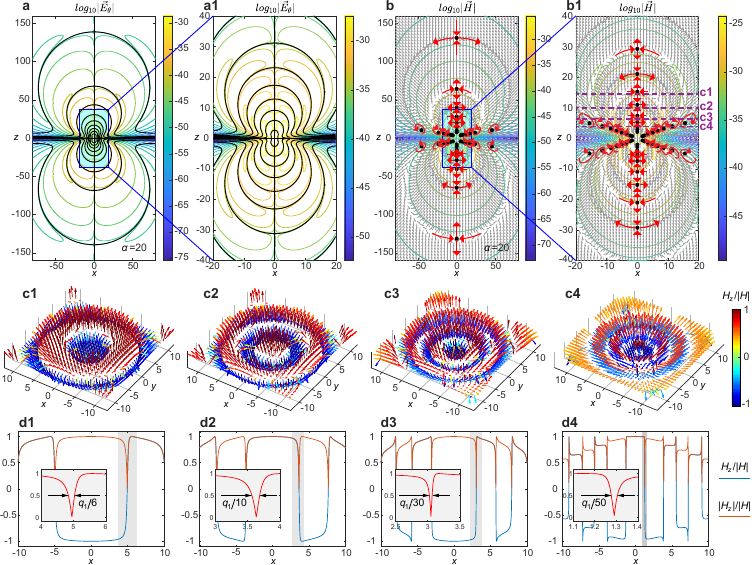}
	\caption{\textbf{Fractal-like pattern in electromagnetic field of supertoroidal pulses:} \textbf{a}, The isoline plot of the electric field in the $x$-$z$ plane for the STLP of $\alpha=20$, $\text{Re}[{E}^{(\alpha=20)}_{\theta}(\mathbf{r},t=0)]$, in logarithmic scale. Solid black lines indicate the zeros of the electric field. Panel \textbf{a1} presents the zoom-in of the region highlighted by blue in (a). \textbf{b}, Isoline and arrow plot of the logarithm of the magnetic field in the $x$-$z$ plane for the STLP of $\alpha=20$. Black dots indicate the zeros of the magnetic field with red arrows correspondingly marking the saddle or vortex style of the vector singularities. Panel \textbf{b1} presents the zoom-in of the region highlighted by blue in (b). \textbf{Fine-scale features of skyrmionic structures:} \textbf{c1-c4}, The skyrmionic distributions of magnetic field at several transverse planes marked by dashed lines \textbf{c1}-\textbf{c4} in (b1). \textbf{d1-d4}, the distribution of normalized magnetic field and its absolute value versus $x$ for the skyrmionic structures in (c1-c4). Insets illustrate the fine-scale features at the regions highlighted by gray bands. Unit for coordinates: $q_1$.} 
	\label{f5}
\end{figure*}

\vspace{0.2cm}
\noindent\textbf{Electric field singularities} --
Figure~\ref{f2} comparatively shows the instantaneous electric fields for the TE single-cycle fundamental TLP and STLP ($\alpha=5$) with $q_2=20q_1$ at the focus ($t=0$). In all cases, the electric field vanishes on the $z$-axis ($r=0$; see the vertical solid black lines in Figs.~\ref{f2}\textbf{a} and \ref{f2}\textbf{b}) owing to the azimuthal polarization and also in the $z=0$ plane (see the horizontal solid black lines in Figs.~\ref{f2}\textbf{a} and \ref{f2}\textbf{b}) due to the odd symmetry of Eqs. Eqs.~(\ref{Ef},\ref{Egf}) with respect to $z$. For the fundamental TLP, the electric field vanishes on two spherical shells (indicated by the solid circles in Fig.~\ref{f2}\textbf{a}) on the positive and negative $z$-axis, respectively. This behavior can be more clearly observed in the transverse distributions at three different propagation distances, $z=q_1,5q_1,35q_1$, of Figs.~\ref{f2}\textbf{a1}-\ref{f2}\textbf{a3}. In accordance to Figs.~\ref{f2}\textbf{a1} and \ref{f2}\textbf{a3}, the electric field  at $z$ positions close to and away from the center of the pulse (cross-sections \textbf{a1} and \textbf{a3}) rotates counter-clockwise forming a vortex around the center singularity along the propagation axis. However, at a distance of $z=5q_1$ from its center (Figs.~\ref{f2}\textbf{a2}), the electric field vanishes on a circular boundary, which corresponds to a spherical region inside which the electric field is oriented along the clockwise direction, whereas outside this region the electric field remains oriented in the opposite (counter-clockwise) direction (see cross-section \textbf{a3}). For the STLP case, a more complex matryoshka-like structure emerges with multiple nested singularity shells, replacing the single shell of the fundamental TLP, as Fig.~\ref{f2}\textbf{b} shows. The electric field configuration close the singularity shells can be examined in detail at transverse planes at $z=q_1,5q_1,35q_1$ (Figs.~\ref{f2}\textbf{b1-b3}). In this case, at transverse planes close to $z=0$, the electric field changes orientation from counter-clockwise close to $r=0$ to clockwise away from the $z$-axis (see Fig.~\ref{f2}\textbf{b1}). On the other hand, on transverse planes close to $z=5q_1$ (see Fig.~\ref{f2}\textbf{b2}), two singular rings (corresponding to the two singular shells) emerge as a cross-sections of the multi-layer singularity shell structure, separating space in three different regions, in which the electric field direction alternates between counter-clockwise ($r/q_1<7$), to clockwise ($7<r/q_1<15$), and again to counter-clockwise ($r/q_1>15$). 

In general, the pulse of higher order of $\alpha$ is accompanied by a more complex multi-layer singular-shell structure, see the dynamic evolution versus the order index in Video~3 in Supplementary Materials. Although the above results of electric fields are instantaneous at $t=0$, we note that the multi layer shall structure propagation of supertoroidal light pulse is retained during propagation, see such dynamic process in Video~4 in Supplementary Materials.

\vspace{0.2cm}
\noindent\textbf{Magnetic field singularities} --
The magnetic field of STLPs has both radial and longitudinal components,  $\mathbf{H}=H_r\hat{\mathbf{r}}+H_z\hat{\mathbf{z}}$, which lead to a topological structure more complex than the one exhibited by the electric field. Figure~\ref{f3} comparatively shows the instantaneous magnetic fields for the TLP and the STLP of $\alpha=5$. For the fundamental TLP (Fig.~\ref{f3}\textbf{a}), the magnetic field has ten different vector singularities on the $x$-$z$ plane, including four saddle points [the longitudinal field component pointing towards (away from) and the radial component away from (toward) the singularity] on \textit{z}-axis and six vortex rings [the surrounding vector distribution forming a vortex loop] away from the \textit{z}-axis. We note that we only consider the singularities existed at an area containing $99.9\%$ of the energy of the pulse. While the singularity existed at the region far away from the pulse center with nearly zero energy can be neglected. A zoom-in of the field structure around these singularities can be seen in Fig.~\ref{f3}\textbf{a1}. For the three off-axis singularities located at the $x>0$ half space, two of them (at $z>0$ and $z<0$) are accompanied by counter-clockwise rotating vortices, whereas the third one (at $z=0$) by a clockwise rotating vortex. Owing to the cylindrical symmetry of the pulse, the off-axis singularities correspond in essence to singularity rings. Such an example is shown in Fig.~\ref{f3}\textbf{a2}, which presents the magnetic field on the transverse plane at $z=0$. Here, the magnetic field points toward the positive (negative) $z$-axis inside (outside) the circular region resulting in the formation of a toroidal vortex winding around the singularity ring. For the STLP (Fig.~\ref{f3}\textbf{b}), more vector singularities are unveiled in the magnetic field with six saddle points on \textit{z}-axis and six off-axis singularities. A zoom-in of the field structure around these singularities can be seen in Fig.~\ref{f3}\textbf{b1}. The orientation of the magnetic field around the on-axis saddle points is alternating between ``longitudinal-toward radial-outward point'' and ``adial-toward longitudinal-outward'', similarly to the on-axis singularities of the TLP. Moreover, the off-axis singularities at $z=0$ become now saddle points contributing to the singularity ring in the  $z=0$ plane. The remaining off-axis singularities are accompanied by clockwise and counterclockwise magnetic field configurations at $x>0$ and $x<0$, respectively as shown Fig.~\ref{f3}\textbf{b2}.

\vspace{0.2cm}
\noindent\textbf{Skyrmionic structure in magnetic field} -- A topological feature of particular interest here is the skyrmionic structure observed in the magnetic field configuration of STLPs. The skyrmion is a topologically protected quasiparticle in condensed matter with a hedgehog-like vectorial field, that gradually changes orientation as one moves away from the skyrmion centre~\cite{fert2017magnetic,yu2010real,nagaosa2013topological}. Recently skyrmion-like configuations have been reported in electromagnetism, including skyrmion modes in surface plasmon polaritons~\cite{tsesses2018optical} and the spin field of focused beams~\cite{du2019deep,gao2020paraxial}. Here we observe the skymrion field configurations in the magnetic field of propagating STLPs. 

The topological properties of a skyrmionic configuration can be characterized by the skyrmion number $s$, which can be separated into a polarity $p$ and vorticity number $m$~\cite{nagaosa2013topological}. The polarity represents the direction of the vector field, down (up) at $r=0$ and up (down) at $r\to\infty$ for $p=1$ ($p=-1$), the vorticity controls the distribution of the transverse field components, and another initial phase $\gamma$ should be added for determining the helical vector distribution, see \textbf{Methods} for details. For the $m=1$ skyrmion, the cases of $\gamma=0$ and $\gamma=\pi$ are classified as N\'eel-type, and the cases of $\gamma=\pm\pi/2$ are classified as Bloch-type. The case for $m=-1$ is classified as anti-skyrmion. 

Here the vector forming skyrmionic structure is defined by the normalized magnetic field $\mathbf{H}=\mathbf{H}/|\mathbf{H}|$ of the STLP. Two examples of two skyrmionic structures in the fundamental TLP are shown in Figs.~\ref{f3}\textbf{c1} ($p=m=1,\gamma=\pi$) and \ref{f3}\textbf{c2} ($p=m=1,\gamma=0$) occurres at the two transverse planes marked by purple dashed lines \textbf{c1} and \textbf{c2}, which are both N\'eel-type skyrmionic structures, where the vector changes its direction from ``down'' at the centre to ``up'' away from the centre. In the case of the STLPs with more complex topology, it is possible to observe more skyrmionic structures. The STLP pulse ($\alpha=5$) exhibits not only the clockwise ($p=m=1,\gamma=\pi$) and counter-clockwise ($p=m=1,\gamma=0$) N\'eel-type skyrmionic structures (Fig.~\ref{f3}\textbf{c3} and \ref{f3}\textbf{c4}), but also those with $p=-1,m=1,\gamma=\pi$ and $p=-1,m=1,\gamma=0$, in Fig.~\ref{f3}\textbf{c6}-\textbf{c6}. 

In general, as the value of $\alpha$ increases, toroidal pulses show an increasingly complex magnetic field pattern with skyrmionic structures of multiple types, see Video~3 in Supplementary Materials. We also note that the topology of the STLP is maintained during propagation, see Video~4 in Supplementary Materials.

\vspace{0.2cm}
\noindent\textbf{Energy backflow and Poynting vector singularities} --
The singularities of the electric and magnetic fields are linked to the  complex topological behavior for the energy flow as represented by the Poynting vector $\mathbf{S}=\mathbf{E}\times\mathbf{H}$. An interesting effect for the fundamental TLP is the presence of energy backflow: the Poynting vector at certain regions is oriented against the prorogation direction (blue arrows in Fig.~\ref{f4}\textbf{a})~\cite{zdagkas2019singularities}. Such energy backflow effects have been predicted and discussed in the context of singular superpositions of waves~\cite{berry2000making,berry2010quantum}, superoscillatory light fields~\cite{yuan2019detecting,yuan2019plasmonics}, and plasmonic nanostructures ~\cite{bashevoy2005optical}. The Poynting vector map reveals a complex multi-layer energy backflow structure, as shown in Fig.~\ref{f4}\textbf{b}. The energy flow vanishes at the positions of the electric and magnetic singularities and inherits their multi-layer matryoshka-like structure. Poynting vector vanishes at $z=0$ plane, along the $z$-axis, and on the dual-layer matryoshka-like singular shells (marked by the black bold lines in Fig.~\ref{f4}\textbf{b}). Importantly, energy backflow occurs at areas of relatively low energy density, and, hence, STLP as a whole still propagates forward. For the temporal evolution of the energy flow of the pulse see Videos~3 \& 4 in Supplementary Materials.

\vspace{0.2cm}
\noindent\textbf{Fractal patterns hidden in electromagnetic fields} --
As the order $\alpha$ of the pulse increases (see Video~3 in Supplementary Material), the topological features of the STLP appear to be organized in a hierarchical, fractal-like fashion. A characteristic case of the STLP of $\alpha=20$ is presented in (Fig.~\ref{f5}). For the electric field, the matryoshka-like singular shells involve an increasing number of layers as one examines the pulse at finer length scales, forming a self-similar pattern that seems infinitely repeated. For the magnetic field, the saddle and vortex points are distributed along the propagation axis and in two planes crossing the pulse centre, respectively. The distribution of singularities becomes increasingly dense as one approaches the centre of the pulse, resulting in a self-similar pattern. A similar pattern can be seen for the Poynting vector map (see Videos~3 \& 4 in Supplementary Materials).


\vspace{0.2cm}
\noindent\textbf{Fine-scale features of skyrmionic structures} -- The fractal-like pattern of vectorial magnetic field of a high-order STLPs results in skyrmionic configurations with with features changing much faster than the effective wavelength $q_1$. Figures~\ref{f5}c1-c4 show the four skyrmionic structures of the high-order STLP ($\alpha=20$) at the four transverse planes marked by the dashed lines in Fig.~\ref{f5}b1 at positions of $z/q_1=14,10,6,3.5$, correspondingly. The four skyrmionic structures have two different topologies with topological numbers of $(p,m,\gamma)=(1,1,\pi)$, for c1 and c3, and $(-1,1,0)$, for c2 and c4. In addition, they exhibit an effect of ``spin reversal'', where the number reversals is given by $\Bar{p}=\frac{1}{2\pi}[\beta(r)]_{r=0}^{r=\infty}$, e.g. $\Bar{p}=1,2,3,4$ for the skyrmionic structures in Figs.~\ref{f5}c1-c4, respectively. Each reversal corresponds to a sign change of $H_z$, which takes place over areas much smaller than the effective wavelength of the pulse ($q_1$).
The full width at half maximum of these areas for the four skyrmionic structures is 1/6, 1/10, 1/30, 1/50 of the effective wavelength, respectively. Conclusively, the sign reversals become increasingly rapid in transverse planes closer to the pulse center ($z=0$), see Figs.~\ref{f5}d1-d4. Similarly, increasing the value of $\alpha$ leads to increasingly sharper singularities.
Notably, in contrast to the fundamental TLP, the skyrmionic configurations in STLP occur at areas of higher energy density, and thus we expect that they could be observed experimentally. 

\begin{figure}[t!]
	\centering
	\includegraphics[width=1\linewidth]{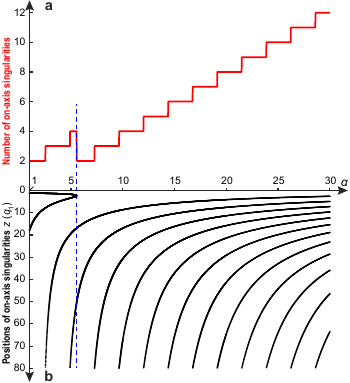}
	\caption{\textbf{Evolution of on-axis singularity distribution versus supertoroidal order:} \textbf{a,b}, The number (\textbf{a}) and positions (\textbf{b}) of on-axis saddle-singularities of the magnetic field of the STLP ($q_2=20q_1$) within the range of $z\in[0,80q_1]$, versus $\alpha$. The blue dashed line marks where the number of singularities decreases.} 
	\label{f6}
\end{figure}

The topological structure of the STLP is directly related to the distribution of on-axis saddle-points in its magnetic field. Indeed, the latter mark the intersection of the $E$-field singular shells with the $z$-axis, which in turn results in the emergence of different skyrmionic magnetic field patterns (see Fig.~\ref{f5} and Videos 3 \& 4). The number and position of on-axis magnetic field saddle points is defined by the supertoroidal parameter $\alpha$. This is illustrated in Fig.~\ref{f6}a, where we plot the number of on-axis $H$-field singularities as a function of alpha for a STLP with $q_2=20q_1$. The number of singularities is generally increasing with increasing $\alpha$ apart for values around $\alpha=5.6$ (marked by blue dashed line in Fig.~\ref{f6}). Moreover, the number of singularities increases in a ladder-like fashion, where only specific values of alpha lead to additional singularities. The origin of this behaviour can be traced to changes in the pulse structure as $\alpha$ increases (see Fig.~\ref{f6}b). For specific values of $\alpha$, additional singularities appear away from the pulse center ($z=0$) and then move slowly towards it. On the other hand, the irregular behavior at $\alpha=5.6$ is a result of two singularities disappearing (see blue dashed line in Fig.~\ref{f6}b). 

\vspace{0.2cm}
\noindent\textit{\textbf{Discussion}}

STLPs exhibit complex and unique topological structure. The electric field exhibits a matryoshka-like configuration of singularity shells, which divide the STLP into ``nested'' regions with opposite azimuthal polarization. The maagnetic field exhibits multi-texture skyrmionic structures at various transverse planes of a single pulse, related to the distribution of multiple saddle and vortex singularities. The instantaneous Poynting vector field exhibits multiple singularities with regions of energy backflow. The singularities of the STLP appear to be hierarchically organized resulting in self-similar, fractal-like patterns for higher-order pulses.

In conclusion, to the best of out knowledge, STLPs are so far the only known example of free-space propagating skyrmionic field configurations. Their structure contains sharp singularities of interest for super-resolution metrology and microscopy, optical information encoding, optical trapping and particle acceleration.

\vspace{0.2cm}
\noindent\textit{\textbf{Methods}}

\noindent\textbf{Solving the supertoroidal pulses} -- The first step is to solve the scalar generating function $f(\mathbf{r},t)$ that satisfies the wave equation $( \nabla^2-\frac{1}{{{c}^{2}}}\frac{{{\partial }^{2}}}{\partial {{t}^{2}}})f\left( \mathbf{r},t \right)=0$,
where $\mathbf{r}=(r,\theta,z)$ are cylindrical coordinates, $t$ is time, $c=1/\sqrt{\varepsilon_0\mu_0}$ is the speed of light, and the $\varepsilon_0$ and $\mu_0$ are the permittivity and permeability of medium. The exact solution of $f(\mathbf{r},t)$ can be given by the modified power spectrum method as $f(\mathbf{r},t)={f_0}/\left[{( {{q}_{1}}+i\tau){{( s+{{q}_{2}})}^{\alpha }}}\right]$, where $f_0$ is a normalizing constant, $s=r^2/(q_1+i\tau)-i\sigma$, $\tau=z-ct$, $\sigma=z+ct$, $q_1$ and $q_2$ are parameters with dimensions of length and act as effective wavelength and Rayleigh range under the paraxial limit, while $\alpha$ is a real dimensionless parameter that must satisfy $\alpha\ge1$ to ensure finite energy solutions. The next step is constructing the Hertz potential. For fulfilling the toroidal symmetric and azimuthally polarized structure, the Hertz potential should be constructed as $\mathbf{A}(\mathbf{r},t)=\mu_0\bm{\nabla} \times\mathbf{\hat{z}}f(\mathbf{r},t)$. Then, the exact solutions of solutions of transverse electric (TE) and transverse magnetic (TM) modes are readily obtained by using Hertz potential. The electromagnetic fields for the TE solution can be derived by the potential as $\mathbf{E}(\mathbf{r},t)=-{{\mu }_{0}}\frac{\partial }{\partial t}\bm{\nabla} \times \mathbf{A}$ and $\mathbf{H}(\mathbf{r},t)=\bm{\nabla} \times(\bm{\nabla}\times \mathbf{A})$~\cite{ziolkowski1989localized,hellwarth1996focused}, see Supplementary Information for more detailed derivations.

\vspace{0.2cm}
\noindent\textbf{Characterizing topology of skyrmion} -- The topological properties of a skyrmionic configuration can be characterized by the skyrmion number defined by~\cite{nagaosa2013topological}:
\begin{equation}
    s=\frac{1}{4\pi }\iint{\mathbf{n}\cdot \left( \frac{\partial \mathbf{n}}{\partial x}\times \frac{\partial \mathbf{n}}{\partial y} \right)}\text{d}x\text{d}y
\end{equation}
that is an integer counting how many times the vector $\mathbf{n}(x,y)= \mathbf{n}(r\cos\theta, r\sin\theta)$ wraps around the unit sphere. For mapping to the unit sphere, the vector can be given by $\mathbf{n}=(\cos\alpha (\theta )\sin\beta (r),\sin\alpha (\theta )\sin\beta (r),\cos\beta (r))$. Also, The skyrmion number can be separated into two integers: 
\begin{align}
\nonumber
s=&\frac{1}{4\pi }\int_{0}^{\infty }{\text{d}r}\int_{0}^{2\pi }{\text{d}\varphi }\frac{\text{d}\beta (r)}{\text{d}r}\frac{\text{d}\alpha (\theta )}{\text{d}\theta }\sin \beta (r)\\
=&\frac{1}{4\pi }[\cos\beta (r)]_{r=0}^{r=\infty }[\alpha (\theta )]_{\theta =0}^{\theta =2\pi }=p\cdot m
\end{align}
the polarity, $p=\frac{1}{2}[\cos\beta (r)]_{r=0}^{r=\infty }$, represents the direction of the vector field, down (up) at $r=0$ and up (down) at $r\to\infty$ for $p=1$ ($p=-1$). The vorticity number, $m=\frac{1}{2\pi }[\alpha (\theta )]_{\theta =0}^{\theta =2\pi }$, controls the distribution of the transverse field components. In the case of a helical distribution, an initial phase $\gamma$ should be added, $\alpha (\theta )=m\theta +\gamma$. For the $m=1$ skyrmion, the cases of $\gamma=0$ and $\gamma=\pi$ are classified as N\'eel-type, and the cases of $\gamma=\pm\pi/2$ are classified as Bloch-type. The case for $m=-1$ is classified as anti-skyrmion. 

\vspace{0.2cm}
\textit{\noindent\textbf{Acknowledgements}}

The authors acknowledge the supports of the MOE Singapore (MOE2016-T3-1-006), the UKs Engineering and Physical Sciences Research Council (grant EP/M009122/1, Funder Id: http://dx.doi.org/10.13039/501100000266), the European Research Council (Advanced grant FLEET-786851, Funder Id: http://dx.doi.org/10.13039/501100000781), and the Defense Advanced Research Projects Agency (DARPA) under the Nascent Light Matter Interactions program.

\vspace{0.2cm}
\textit{\textbf{Data availability}}
The data from this paper can be obtained from the University of Southampton ePrints research repository https://doi.org/xx.xxxx/SOTON/xxxxx.

{$^*$y.shen@soton.ac.uk}



\bibliography{apssamp}
%

\end{document}